\documentclass[sigconf,screen]{acmart}
\AtBeginDocument{%
  \providecommand\BibTeX{{%
    \normalfont B\kern-0.5em{\scshape i\kern-0.25em b}\kern-0.8em\TeX}}}
  
\usepackage{amsfonts}
\usepackage{algorithmic}
\usepackage{textcomp}
\def\BibTeX{{\rm B\kern-.05em{\sc i\kern-.025em b}\kern-.08em
    T\kern-.1667em\lower.7ex\hbox{E}\kern-.125emX}}

\usepackage{algorithmic}
\usepackage{color}
\usepackage{url}
\usepackage{enumitem}
\usepackage{multirow}
\usepackage[ruled, vlined, linesnumbered]{algorithm2e}
\usepackage{subcaption}
\SetAlCapNameFnt{\scriptsize}
\SetAlCapFnt{\small}

\usepackage{listings}
\usepackage{verbatim}

\usepackage{xspace}

\usepackage{soul}

\newcommand*{\eg}{e.g., }
\newcommand*{\ie}{i.e., }

\DeclareFontFamily{OT1}{mathc}{}
\DeclareFontShape{OT1}{mathc}{m}{it}{<-> mathc10}{}
\DeclareMathAlphabet{\mathabxcal}{OT1}{mathc}{m}{it}

\newcommand{\mypara}[1]{\vspace{0.2em}\noindent\textit{\textbf{#1}} }

\usepackage{colortbl}

\usepackage{pifont}

\definecolor{mygreen}{HTML}{02818a}
\definecolor{mypurple}{HTML}{8f3c8c}

\newcommand{\tool}{\textsc{Basic}}

\newcommand{\keepnotes}{true}

\ifthenelse{\isundefined{\keepnotes}}{
\newcommand{\mytodogreen}[1]{\textcolor{mygreen}{\ding{46}~{\sf}~#1}}
\newcommand{\yibiao}[1]{\mytodogreen{[yibiao: #1]}}
\newcommand{\civi}[1]{\textcolor{red}{[Ming: #1]}}
\newcommand{\yj}[1]{\textcolor{blue}{[yj: #1]}}
\newcommand{\del}[1]{\textcolor{red}{\sout{#1}}}
}{
\newcommand{\mytodogreen}[1]{}
\newcommand{\civi}[1]{}
\newcommand{\yj}[1]{}
\newcommand{\yibiao}[1]{}
\newcommand{\del}[1]{}
}
    
\setcopyright{none}
\setcopyright{acmlicensed}

\author{Yibiao Yang, Qingyang Li, Maolin Sun, Jiangchang Wu, Yuming Zhou}

\affiliation{%
  \institution{State Key Laboratory for Novel Software Technology, Nanjing University, Nanjing, China\\yangyibiao@nju.edu.cn, \{liqingyang, merlin, jiangchangwu\}@smail.nju.edu.cn, zhouyuming@nju.edu.cn}
  \country{}
}

\title{Using a Sledgehammer to Crack a Nut? Revisiting Automated Compiler Fault Isolation}

\begin{document}

\begin{abstract}
  \textbf{\textit{Background:}} Compilers are fundamental to software development, translating high-level source code into executable software systems. 
Faults in compilers can have severe consequences and thus effective localization and resolution of compiler bugs are crucial. 
\textbf{\textit{Problem:}} In practice, developers often examine version history to identify and investigate bug-inducing commit (BIC) for fixing bugs. However, while numerous sophisticated Spectrum-Based Fault Localization (SBFL) techniques have been proposed for compiler fault isolation, their effectiveness has not been evaluated against the BIC-based strategies widely adopted in practice. 
\textbf{\textit{Objective:}} This study aims to bridge this gap by directly comparing a BIC-based strategy, \tool, with representative SBFL techniques in the context of compiler fault localization.
The BIC-based strategy closely aligns with common developer practices, as it directly identifies the BIC and treats the files modified in that commit as faulty candidates. 
\textbf{\textit{Method:}} The \tool~identifies the most recent \textit{good release} and earliest \textit{bad release}, and then employs a binary search to pinpoint the bug-inducing commit. All files modified in the identified commit are flagged as potentially faulty. 
We rigorously compare \tool~against SBFL-based techniques using a benchmark consisting of 60 GCC bugs and 60 LLVM bugs. 
\textbf{\textit{Result:}} Our analysis reveals that \tool~performs comparably to, and in many cases outperforms, state-of-the-art SBFL-based techniques, particularly on the critical Top-1 and Top-5 ranking metrics. 
\textbf{\textit{Conclusion:}} 
This study provides new insights into the practical effectiveness of SBFL-based techniques in real-world compiler debugging scenarios. 
We recommend that future research adopt \tool~as a baseline when developing and evaluating new compiler fault isolation methods.

\end{abstract}

\ccsdesc[300]{Software and its engineering~Compilers}
\ccsdesc[500]{Software and its engineering~Software testing and debugging}

\keywords{Compiler, Fault Isolation, Bug inducing commit, SBFL}

\maketitle

\section{Introduction}
\label{sec:intro}
Compilers are essential components in software development, responsible for translating high-level programming languages into low-level, executable machine code~\cite{alfred2007compilers}. 
Due to their scale and complexity, however, compilers are prone to faults that can have severe downstream consequences~\cite{yang2011finding}. 
Such faults often manifest as ``silent'' miscompilations, where incorrect binaries are produced without obvious failures, often misleading developers into attributing the observed errors to defects in their own code rather than the underlying compiler. 
As a result, developers may incur substantial and unnecessary debugging effort. 

A substantial body of prior research has focused on testing techniques for compiler toolchains, including differential testing, metamorphic testing, and fuzzing~\cite{csmith-pldi,emi-pldi,yarpgen,yarpgenv2,atlas-atc,clozemaster,creal-pldi,skl-pldi,DBLP:conf/sosp/Li00S23,devil-asplos,c2v-icse,cod-ase,decov-fse}. These techniques have been highly effective in uncovering large numbers of real-world compiler bugs that require prompt resolution.
Consequently, effective techniques for isolating and resolving these compiler faults are crucial for ensuring the reliability and correctness of downstream software systems~\cite{whalley1994automatic}. 

Identifying the specific source files responsible for compiler bugs remains challenging due to compiler's size and complexity~\cite{chen2019compiler}. 
To address this challenge, prior studies have proposed a variety of automated Spectrum-Based Fault Localization (SBFL) techniques for locating compiler faults.
These approaches construct execution spectra by running the compiler on both failing and passing inputs, where passing inputs are typically obtained either by mutating failing programs~\cite{chen2019compiler,chen2020enhanced,DBLP:journals/tse/TuZJYLJ24} or by systematically adjusting compilation options~\cite{ODFL-TSE}. 
Based on the collected coverage information, SBFL formulas compute suspiciousness scores for program elements, such as source files and function, to guide fault localization~\cite{abreu2007accuracy,wong2013dstar}. 

Although SBFL-based techniques have demonstrated promising results for compiler fault localization~\cite{chen2019compiler,chen2020enhanced,ODFL-TSE}, their effectiveness has rarely been examined against straightforward debugging strategies commonly employed by compiler developers in practice.
In practice, developers frequently examine version histories to identify bug-inducing commits (BICs), a code change commit that introduces a bug absent in earlier versions, and analyze the associated code changes to diagnose compiler faults~\cite{gccreghunt,llvmbisect}. 
Modern version control systems support fine-grained tracking of source code evolution. 
For instance, LLVM documentation explicitly recommends using \texttt{git bisect} to identify faulty revisions~\cite{llvmbisect}, while GCC documentation outlines procedures for identifying the release in which a bug first appears~\cite{gccreghunt}. 
In practice, developers perform a binary search over the commit history and then inspect the files modified in the identified BIC to localize compiler faults~\cite{gccreghunt,llvmbisect}. 

While BIC identification has been extensively studied in the context of defect prediction, bug-fix pattern mining, and automated program repair~\cite{DBLP:conf/icse/AnHKY23,10711218,4019564,10.1145/3377816.3381743,8987574,10.1007/s10664-008-9077-5}, these efforts do not directly leverage BICs for fault localization.
Among existing fault localization techniques, HSFL~\cite{DBLP:journals/tse/WenCTWHHC21} is the only approach that explicitly incorporates BIC information. However, HSFL leverages historical BIC data solely to adjust suspiciousness scores, rather than treating the identified BIC itself as the fault localization results. 
More recently, Zhou et al.~\cite{ZhouXuSun2025} utilized BICs to support compiler bug deduplication, further highlighting the practical value of revision history in compiler debugging.
Nevertheless, despite the central role of BIC identification in real-world debugging workflows, prior studies have neither systematically examined BIC identification as a standalone, end-to-end fault localization strategy nor rigorously compared it against SBFL-based techniques. 

In this work, we introduce \tool, a practitioner-aligned compiler fault localization approach that closely mirrors real-world debugging workflows.
Specifically, \tool~first determines the most recent \textit{good} release and the earliest \textit{bad} release, and then applies binary search to locate the BIC between them. 
Unlike SBFL-based approaches that rely on execution spectra, \tool~localizes faults through release-level comparison and bisection, treating the files modified in the identified commit as faulty candidates.
We conduct a comprehensive empirical study comparing \tool~with representative SBFL-based compiler fault localization techniques on widely used GCC and LLVM benchmarks.

Our findings reveal the surprising effectiveness of \tool~in real-world scenarios. Using benchmarks comprising 60 GCC bugs and 60 LLVM bugs, we find that \tool~achieves performance comparable to, and often exceeding, that of state-of-the-art SBFL-based techniques, including DiWi~\cite{chen2019compiler}, RecBi~\cite{chen2020enhanced}, LLM4CBI~\cite{DBLP:journals/tse/TuZJYLJ24}, \textsc{Odfl}~\cite{ODFL-TSE}, and HSFL~\cite{DBLP:journals/tse/WenCTWHHC21}, in accurately identifying faulty files.
These results expose important limitations of current SBFL-based approaches and suggest that BIC-based localization consitutes a strong and practical baseline for compiler fault localization research. 

\mypara{Contributions.} Our main contributions are as follows:

\begin{itemize}[leftmargin=*]
    \item \textbf{Practitioner-Centric Perspective.} We revisit compiler fault localization from the perspective of real-world debugging practice and formalize BIC identification via bisection as a standalone, end-to-end fault localization strategy for compilers. 
    \item \textbf{Technique.} We propose \tool, a simple yet effective approach that directly identifies the BIC and treats the files modified in that commit as faulty candidates, closely align with the workflow commonly adopted by compiler developers.
    \item \textbf{Empirical Evaluation.} We conduct a comprehensive empirical study that systematically compares \tool~with existing representative SBFL-based compiler fault localization techniques using consistent datasets and evaluation settings. 
    \item \textbf{Implications.} Our results demonstrate that \tool~performs comparable to, and in many cases outperforms, state-of-the-art SBFL-based compiler fault localization techniques. Beyond compilers, \tool~is applicable to debugging a wide range of version-controlled software systems, as it relies solely on commit analysis and thus offers strong generality.
    These findings suggest that: (1) new fault localization techniques, especially for compilers, should demonstrate improvements not only over traditional SBFL methods but also over practical BIC-based baselines such as \tool; and (2) hybrid strategies that integrate SBFL with BIC information may offer further benefits and warrant investigation.
\end{itemize}

\mypara{Organization.} This paper is structured as follows: Section~\ref{sec:background} provides an overview of compiler fault localization, including examples that illustrate the motivation for our study. Section~\ref{sec:approach} details the employed methodology.
Section~\ref{sec:settings} describes the experimental setup for our comparative study. Section~\ref{sec:evaluation} presents the results and analysis of our investigation, followed by a discussion in Section~\ref{sec:validity}. Section~\ref{sec:related-work} reviews related works. Finally, Section~\ref{sec:conclusion} summarizes our findings and suggests potential avenues for future research.

\section{Background \& Motivation}
\label{sec:background}
In this section, we provide a comprehensive background on Spectrum-Based Fault Localization (SBFL) techniques used in compiler fault localization, as well as the code-change-based Version Control System (VCS) employed in modern software development. We then present the motivation behind our approach with illustrative examples of real-world bugs in GCC and LLVM compilers.

\lstdefinestyle{coloredbash}{
    language=bash,
    basicstyle=\small\ttfamily,
    moredelim=[s][\color{green}]{@}{@}, 
    moredelim=[s][\color{red}]{!}{!} 
}

\begin{figure*}[t]
  \begin{minipage}{0.45\textwidth}
    \begin{subfigure}{\textwidth}
      \centering
      \begin{lstlisting}[
        language=c,
        basicstyle=\footnotesize\ttfamily,
        frame=single,
        numbers=left,
        backgroundcolor=\color{gray!5}
      ]
int a[6], b, c = 1, d;
short e;

void fn1 (int p) { b = a[p]; }

int main () {
  a[0] = 1;
  if (c) e--;
  d = e;
  long long f = e;
  fn1 ((f >> 56) & 1);
  printf ("%
  return 0;
}
      \end{lstlisting}
      \caption{Bug-triggering test program}\label{fig:gcc-bug-59747-triggering-test-program}
    \end{subfigure}
  \end{minipage}%
\hspace{2em}
\begin{minipage}{0.45\textwidth} 
\begin{subfigure}{\textwidth}
\centering
\begin{lstlisting}[
    language=bash,
    basicstyle=\footnotesize\ttfamily\color{black},
    backgroundcolor=\color{gray!5},
    frame=single,
    escapeinside={(*@}{@*)}, 
    linewidth=\textwidth
]
$ gcc -v
gcc version 4.9.0 (trunk 206472)
(*@\colorbox{green!15}{\makebox[\dimexpr\linewidth-3pt][l]{\$ gcc -m64 -O1 small.c; a.out}}@*)
(*@\colorbox{green!15}{\makebox[\dimexpr\linewidth-3pt][l]{0}}@*)
(*@\colorbox{red!15}{\makebox[\dimexpr\linewidth-3pt][l]{\$ gcc -m64 -Os small.c; a.out}}@*)
(*@\colorbox{red!15}{\makebox[\dimexpr\linewidth-3pt][l]{1}}@*)
\end{lstlisting}
\caption{The compilation command that triggers the bug}\label{fig:gcc-bug-59747-compile}
\vspace{1em}
\end{subfigure}

    \begin{subfigure}{\textwidth}
      \centering
      \begin{lstlisting}[
        basicstyle=\footnotesize\ttfamily,
        language=bash,
        frame=single,
        backgroundcolor=\color{red!15}
      ]
$ svn diff -r 206417:206418 --summarize \
    | grep -E 'gcc/[A-Za-z\-]+\.c$'
M       gcc/ree.c
      \end{lstlisting}
      \caption{The files modified in the bug-inducing revision of r206418} \label{fig:gcc-bug-59747-revision}
    \end{subfigure}
\end{minipage}
  
\caption{GCC bug \#\href{https://gcc.gnu.org/bugzilla/show_bug.cgi?id=59747}{59747}. When compiling the program with the flag `m64 -O1' using GCC trunk revision r206472, the binary prints `0' as expected. However, when compiling with the flag `m64 -Os', the binary erroneously prints `1'.}
\label{fig:gcc-bug-59747}
\end{figure*}

\lstdefinestyle{coloredbash}{
    language=bash,
    basicstyle=\small\ttfamily,
    moredelim=[s][\color{green}]{@}{@}, 
    moredelim=[s][\color{red}]{!}{!} 
}

\begin{figure*}[t]
  \begin{minipage}{0.45\textwidth}
    \begin{subfigure}{\textwidth}
      \centering
      \begin{lstlisting}[
        language=c,
        basicstyle=\footnotesize\ttfamily,
        frame=single,
        numbers=left,
        backgroundcolor=\color{gray!5}
      ]
  int main() {
    int a;
    for (a = 2; a >= 0; a--)
      ;
    unsigned b = -1 %
    a = b;
    if (a != 0)
      __builtin_abort ();
    return 0;
  }
      \end{lstlisting}
\caption{Bug-triggering test program}\label{fig:llvm-bug-28307-triggering-test-program}
    \end{subfigure}
  \end{minipage}%
\hspace{2em}
\begin{minipage}{0.45\textwidth} 
\begin{subfigure}{\textwidth}
\centering
\begin{lstlisting}[
    language=bash,
    basicstyle=\footnotesize\ttfamily\color{black},
    backgroundcolor=\color{gray!5},
    frame=single,
    escapeinside={(*@}{@*)}, 
    linewidth=\textwidth
]
$ clang -v
clang version 17.0.0 (trunk ab8d7ea)
(*@\colorbox{red!15}{\makebox[\dimexpr\linewidth-3pt][l]{\$ clang -O1 small.c; ./a.out}}@*)
(*@\colorbox{red!15}{\makebox[\dimexpr\linewidth-3pt][l]{Aborted}}@*)
\end{lstlisting}
\caption{The compilation command that triggers the bug}\label{fig:llvm-bug-61312-compile}
\vspace{1em}
\end{subfigure}

    \begin{subfigure}{\textwidth}
      \centering
      \begin{lstlisting}[
        basicstyle=\footnotesize\ttfamily,
        language=bash,
        frame=single,
        backgroundcolor=\color{red!15}
      ]
$ git diff --name-only 15d5c59^ 15d5c59 -- '*.cpp'
llvm/lib/Analysis/InstructionSimplify.cpp
llvm/lib/Transforms/InstCombine/InstCombineCompares.cpp
      \end{lstlisting}
\caption{The files modified in the bug-inducing commit of 15d5c59} \label{fig:llvm-bug-61312-revision}
    \end{subfigure}
\end{minipage}

\caption{LLVM bug \#\href{https://github.com/llvm/llvm-project/issues/61312}{61312} demonstrates an unexpected abortion when executing this program after compiling it using LLVM trunk revision ab8d7ea with the `-O1' compilation flag.}
\label{fig:llvm-bug-28307}
\vspace{-1.0em}
\end{figure*}

\subsection{Background}

\subsubsection{Spectrum-Based Fault Localization}
SBFL is a widely used technique in software debugging, aiming to identify the root cause of software faults or failures~\cite{abreu2009spectrum,tosem2013xie}.
SBFL leverages coverage information from passing and failing test cases to assess the suspiciousness of individual code elements, ranking them by their likelihood of being faulty~\cite{abreu2009spectrum}. 
The fundamental assumption is that program elements executed more frequently by failing test cases are more likely to be the root cause of the fault, whereas those executed more frequently by passing test cases are less likely. 
SBFL enables more effective debugging by prioritizing efforts on the most suspicious code locations. However, SBFL's effectiveness depends heavily on the quality and quantity of the test cases used~\cite{keller2017critical,tosem2013xie}.
Insufficient failing test cases or inadequate representation of the fault in the test cases can compromise accuracy. 
Prior studies~\cite{keller2017critical} have shown that increasing the number of failing test cases can enhance the effectiveness of SBFL, indicating that greater diversity in coverage information proves beneficial. 
The formula commonly used in SBFL to calculate the suspiciousness score of a code element is based on the this equation:
$ Suspiciousness Score = Failed(e) / (Failed(e) + Passed(e)) $,
where $Failed(e)$ represents the number of failing test cases that cover code element $e$, and $Passed(e)$ represents the number of passing test cases that cover code element $e$. The suspiciousness score reflects the likelihood of a code element being faulty, with higher scores indicating higher suspicion. By ranking code elements based on their suspiciousness scores, SBFL provides guidance to developers in identifying potential fault locations for further investigation and debugging.

\subsubsection{Compiler Fault Isolation Techniques}
Within the context of compiler fault localization, various SBFL-based techniques have been developed. 
Chen et al.~\cite{chen2019compiler,chen2020enhanced} proposed generating witness test programs by randomly mutating bug-triggering test programs to isolate compiler optimization faults. 
The bug-triggering test program exposes bugs in compiler, while the witness test program is not expected to trigger the same bug with the same compilation configuration. 
By combining these test programs, SBFL techniques can be applied to locate faulty files in compilers~\cite{abreu2009spectrum, tang2017accuracy}. 
However, this approach has inherent limitations. 
There is no guarantee that the witness test program can always be generated by randomly mutating the bug-triggering test programs. 
The process of generating a witness test program can be time-consuming, requiring numerous mutation attempts. 
Moreover, the localization results obtained from witness test programs may be unreliable due to the random nature of the mutation strategies employed. 
These limitations make it challenging to effectively locate bugs in compilers. 
To address these limitations, Yang et al.~\cite{ODFL-TSE} proposed an alternative approach focusing on searching for passing and failing compilation configurations with minimal difference. 
By replacing test program mutation with low-cost configuration modifications, this approach achieves more accurate and efficient fault localization. 
These techniques utilize coverage information from executed code to isolate and identify faults within compilers. They involve the generation of passing test programs or configurations by modifying bug-triggering test programs or utilizing fine-grained optimization options. By analyzing the coverage spectra of failing and passing compilations, these techniques calculate the suspiciousness for each program element, such as files, functions, or statements within compilers.

\subsubsection{Commit-Based Version Control System}
Version Control Systems (VCS) are critical role in modern software development for managing code changes. 
Developers use VCS to introduce modifications, or commits, which consist of specific code changes and associated message describing the changes. 
The commit-based approach allows for better collaboration among team members, as it enables them to work on different code changes concurrently and integrate them seamlessly. 
The chronological history of commits helps developers track the evolution of the codebase and promotes accountability and transparency. 
When a fault is reported, developers can leverage the commit history to identify the specific changes that may have introduced the issue. 
By pinpointing the relevant commits, developers can narrow down their focus and conduct more targeted investigations into the code changes, leading to expedited fault isolation. 
In the context of fault localization, a bug-inducing commit refers to a specific commit in the VCS that is responsible for introducing a software fault. When a fault is reported, developers can leverage the commit history to identify the precise changes that may have introduced the issue. By locating the relevant bug-inducing commits, developers can narrow their focus and conduct more targeted investigations into the associated code changes. This process expedites the fault isolation process by enabling developers to effectively trace the root cause of the issue.

\begin{figure*}[ht]
\centering
\includegraphics[width=0.6\linewidth]{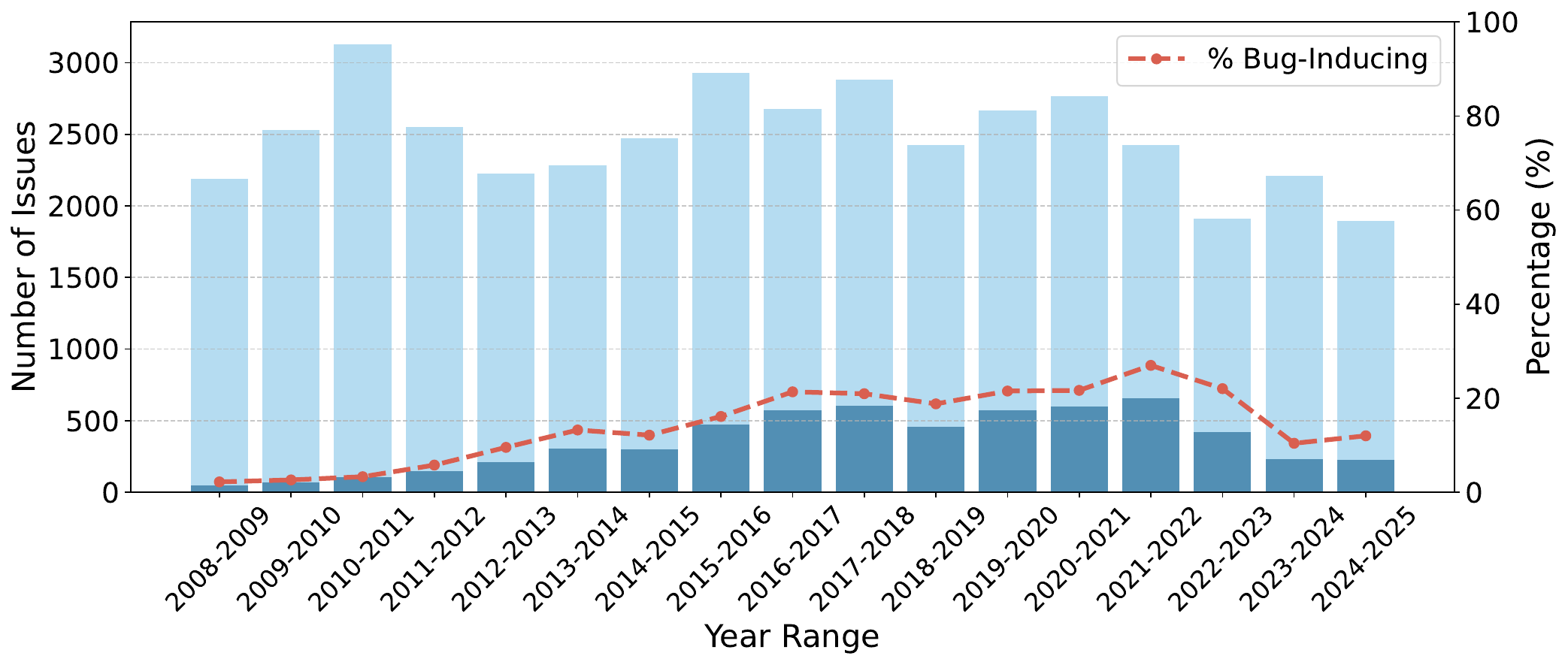}
\vspace{-0.8em}
\caption{Bugzilla issues and BIC-associated issues over time. The blue bars represent the total number of issues, and the deeper blue bars represent the number of issues associated with BIC. The orange line indicates the percentage of BIC-associated issues relative to the total number of issues.}
\label{fig:gcc-bug-inducing-commits}
\vspace{-1em}
\end{figure*}

\subsection{Motivation}\label{sec:motivation}

Our approach is motivated by the need to evaluate and compare the performance of existing SBFL-based compiler fault isolation techniques with a simplified approach commonly employed by developers in practice. While SBFL-based techniques have demonstrated effectiveness in isolating compiler faults, it remains unclear whether they are practically effective when compared to the simplified bug-including commit based approach used by developers. By conducting a comprehensive comparative analysis, we aim to shed light on the effectiveness and limitations of SBFL-based techniques in real-world scenarios and identify areas for improvement. 
In the following, we illustrate our motivation using two fixed issues of the GCC and LLVM compilers. %

\subsubsection{Motivating Example of GCC}

Figure~\ref{fig:gcc-bug-59747} depicts the GCC bug \href{https://gcc.gnu.org/bugzilla/show_bug.cgi?id=59747}{\#59747}, an instance of a wrong code bug within the GCC development version r206472. 
Figure~\ref{fig:gcc-bug-59747-triggering-test-program} presents the test program that triggers the bug, and Figure~\ref{fig:gcc-bug-59747-compile} illustrates the compilation command that exposes it. 
Notably, when compiling the test program using the `-m64 -O0' compilation option with GCC development version r206472, the resulting binary program produces an execution output of 0. 
However, when using the `-m64 -Os' option to compile the same test program, the execution output of the generated binary program yields a value of 1, an unexpected behavior. 
According to the bug tracking system's description page for bug \#59747, 
developer \texttt{Jakub Jelinek} commented on the issue's initiation with r206418. 
Figure~\ref{fig:gcc-bug-59747-revision} showcases the modifications made in revision r206418, obtained through the command ``\texttt{svn diff -r 206417:206418 --summarize \textbar grep -E `gcc/[A-Za-z\textbackslash-]+\textbackslash.c'}.''
This command displays a diff between revisions r206418 and r206417, revealing that the sole modification occurred in the `gcc/ree.c' source file within the GCC source code induced the bug.
\texttt{Jeffrey A. Law}'s comment on the description page indicates that the bug was resolved in revision r206638.
Except a test program of `pr59747.c' adding to the gcc testsuite folder, this bug-fixing revision only modified the `gcc/ree.c' file within the gcc source code folder. 
The modified source file in the bug-fixing commit exactly aligns with the file modified in the bug-inducing commit. 
This illustrative example demonstrates the feasibility of utilizing files modified by bug-inducing commits to identify faulty files within the GCC compiler.

\begin{figure*}[t]
    \centering
\includegraphics[width=0.7\linewidth]{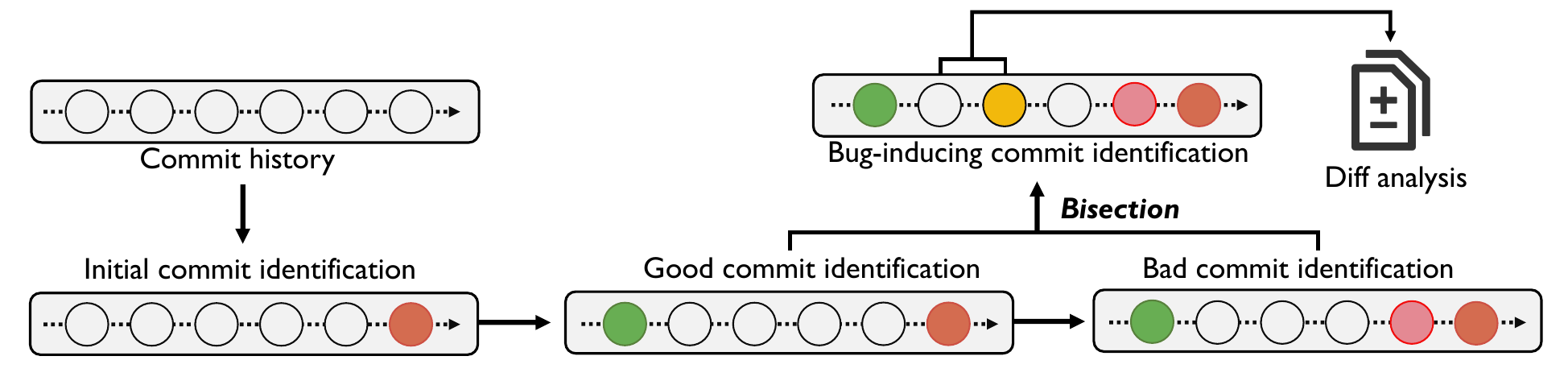}
\vspace{-1em}
	\caption{Overview of \tool. \label{fig:overview}}
 \vspace{-1em}
\end{figure*}

\subsubsection{Motivating Example of LLVM}
Figure~\ref{fig:llvm-bug-28307} displays the LLVM bug \href{https://github.com/llvm/llvm-project/issues/61312}{\#61312}, reported on the LLVM GitHub issue tracker and subsequently fixed by the developers.
This bug caused by incorrect code generation by the LLVM compiler, resulted in unexpected termination of the generated binary when executed.
Figure~\ref{fig:llvm-bug-28307-triggering-test-program} presents the test program that triggers this bug, while Figure~\ref{fig:llvm-bug-61312-compile} illustrates the compilation command that exposed the bug.
After compiling the test program using the `-O1' compilation option with LLVM trunk revision ab8d7ea, the resulting binary program aborted when executed.
However, the program should not terminate as the variable `a' is set to 0, which should not activate the `\texttt{\_\_builtin\_abort()}' function.
Developer \texttt{Nikita Popov} noted that this bug was introduced in revision 15d5c59.
Given this information, we can utilize the command ``\texttt{git diff --name-only 15d5c59\textasciicircum 15d5c59 -- `llvm/lib/*.cpp'}'' to identify the files modified in the bug-inducing commit of 15d5c59 compared to its parent commit (\ie~revision 15d5c59\textasciicircum).
This command reveals that the bug-inducing commit modified the `llvm/lib/Analysis/InstructionSimplify.cpp' and `llvm/lib/Transforms/InstCombine/InstCombineCompares.cpp' source files within the LLVM source code, as indicated in Figure~\ref{fig:llvm-bug-61312-revision}.
The inspection of the bug-fixing commit showed that the modified files were identical to those affected in the bug-inducing commit.
This highlights the importance of examining bug-inducing commits to identify faulty files in the LLVM compiler.

These two examples are not isolated cases, but rather reflect a broader pattern in real-world compiler bug reports. To investigate this trend, we further conducted an empirical analysis of issue tracking systems of GCC and LLVM.
For GCC, we analyzed all resolved bug reports in Bugzilla, over 40,000 in total. Among these, 6,365 reports contain the phrase ``\textit{started with}'' in their discussion, which is a common way for GCC developers to indicate the bug-inducing commit~\cite{gccbug104334,gccbug103961}. This suggests that these developers often use \texttt{git bisect} or similar mechanisms to identify the faulty change. 
Figure~\ref{fig:gcc-bug-inducing-commits} plots the yearly number of bug reports, highlighting those that explicitly mention a bug-inducing commit. 
Before 2010, such references were relatively rare, but their frequency has steadily increased over time. In 2021–2022, nearly 30\% of all bug reports explicitly identified a bug-inducing commit, indicating a growing reliance on this technique. This trend is likely driven by the increasing volume of bug reports, often exceeding 2,000 per year, and the efficiency gains bisection provides in fault isolation. 
Note that our measurement is conservative. Developers sometimes use alternative wording to indicate a bug-inducing commit, and they are not required to document the bug-inducing commit in every report, especially when the debugging process does not involve coordination with others.
As a result, the actual prevalence of bisection is likely higher than what our keyword-based analysis captures.
For LLVM, we performed a similar analysis using its GitHub issue tracker. 
From roughly 60{,}000 issues, we searched for indicative terms such as ``\textit{bisect}'' and ``\textit{introduced in}'' within issue descriptions and comments. We found 1,720 closed issues containing ``\textit{bisect}'' and 1,500 containing ``\textit{introduced in}.'' In most of these cases, developers explicitly referenced the commit that introduced the bug. Although these keywords do not capture all bisection instances, they provide a clear signal that LLVM developers also apply bisection to isolate faults.
Overall, our study shows that both GCC and LLVM developers routinely use bisection in their debugging workflows. This finding grounds our motivation for evaluating a fault localization approach that aligns with this established practice.

\section{Approach}
\label{sec:approach}

This section introduces \tool, a compiler fault isolation technique that accurately identifies bug-inducing changes and considers all modified source code files as potential faulty files within the compiler. 
\tool~consists of five primary steps, as shows in Figure~\ref{fig:overview}. 
We will explain each step in detail in subsequent sections.

\subsection{Framework}

Algorithm~\ref{algo:overall_algo} depicts the detailed workflow of \tool.

\subsubsection{Initial Commits Identification}
The first step of \tool~is to retrieve the commit identification (ID) from the bug report description (line~\ref{algo:info_retrieve}).
Typically, a well-crafted bug report includes details about the specific version of the compiler where the bug manifests. 
This identified commit serves as the starting point for subsequent analysis, where we examine the previous commit to find the bug-inducing changes.

\subsubsection{Rough Range Identification}
Upon retrieving the commit from the bug report, \tool~proceeds to identify the rough ranges of the \textit{bad commit} and the \textit{good commit}.
Concretely, \tool~first aims to identify the oldest major version\footnote{
A major version is defined as a significant release of the compiler, usually marked by a change in the first digit of the version number, \eg~GCC 9.1 versus GCC 10.1.
} of the compiler that still manifests the bug under investigation. 
To accomplish this, \tool~systematically examines each major version to identify the earliest one that still exhibits the bug (lines~\ref{algo:bad_commit:start}-\ref{algo:bad_commit:end}).
The commit associated with this earliest failing major version is then designated as the \textit{bad commit}.
Notably, if no major version reproduces the reported bug, the initial commit is assigned as the \textit{bad commit}.
Accordingly, the \textit{good commit} will be the commit associated with the latest passing major version of the compiler, \ie~the major version preceding the \textit{bad commit} that does not manifest the reported bug (line~\ref{algo:good_commit}).
This preliminary investigation significantly narrows the search space for the subsequent bisection process aimed at identifying the bug-inducing commit, thus enhancing the efficiency of the fault isolation process.

\subsubsection{Fine Range Identification}
Subsequently, \tool~identifies the fine range between the \textit{bad commit} and the
\textit{good commit}. \tool~focus on the minor versions\footnote{Minor versions refer to subsequent releases within a major version, typically marked by a change in the second digit of the version number, e.g., GCC 9.1 versus GCC 9.2.} within the rough range previously identified (lines~\ref{algo:minor:start}-\ref{algo:minor:end}).
Specifically, it first searches for the most recent minor version of the compiler that does not exhibit the reported bug. 
This version should be the one closest to the earliest failing major version within the rough range that does not trigger the bug. 
By iteratively examining the minor versions between the rough range.
\tool~identifies the most recent passing minor version. It then designates the commit of this version as the updated \textit{good commit}.
Next, \tool~examines the minor versions within the rough range to pinpoint the earliest failing minor version.
Through iterative examination, it identifies the commit of the earliest failing minor version and updates the \textit{bad commit} accordingly.
This process narrows down the search space for subsequent steps.

\subsubsection{Bug-inducing Commit Identification}
With the \textit{bad commit} and  \textit{good commit} IDs within the fine range obtained, \tool~employs a binary search strategy to precisely pinpoint the specific bug-inducing commit. 
Binary search, a well-known algorithm, efficiently searches for a specific element in a sorted collection. 
In the context of \tool, the sorted collection represents the commit history between the \textit{bad commit} and \textit{good commit}, ordered by commit time. 
By iteratively narrowing down the search space with binary search, \tool~performs a series of comparisons between midpoint commits until accurately identifying the bug-inducing commit. 
This bisection significantly reduces the search space and accelerates \textit{bug-inducing commit} identification. 

\subsubsection{Differential Analysis}
Once the \textit{bug-inducing commit} has been identified, \tool~conducts a differential analysis between the bug-introducing change and its preceding change. 
The differential analysis aims to identify the specific code files modified by the \textit{bug-inducing commit}. 
Specifically, by comparing the content and structure of the code files before and after the \textit{bug-inducing commit}, \tool~identifies the differences and determines the set of modified files. 
This process can be achieved by utilizing the \texttt{git diff} or \texttt{svn diff} command and the \texttt{grep} command, as exemplified in the illustrative example presented in Section~\ref{sec:motivation}. 
Therefore, the differential analysis enables \tool~to identify the code files that were modified by the \textit{bug-inducing commit} as the faulty files. 
This information is crucial for further investigation and resolution of the bug.

\newcommand\mycommfont[1]{\footnotesize\ttfamily\textcolor{blue}{#1}}
\SetCommentSty{mycommfont}

\setlength{\textfloatsep}{0.5cm}

\begin{algorithm}[t]
\SetKwFunction{FnInfoRetrieve}{InfoRetrieve}
\SetKwFunction{FnCompile}{Compile}
\SetKwFunction{FnGetRelease}{GetBranchPoint}
\SetKwFunction{FnGetReleaseCommit}{GetReleaseCommit}
\SetKwFunction{FnBinarySearch}{BinarySearch}
\SetKwFunction{FnGetMinorReleases}{GetMinorReleases}
\SetKwFunction{FnGetMajorReleases}{GetMajorReleases}
\SetKwFunction{FnGetPrecedingRelease}{GetPrecedingMajorRelease}
\SetKwFunction{FnDate}{ReleaseDate}
\SetKwFunction{FnBisect}{Bisect}
\SetKwFunction{FnDiffAnalysis}{DiffAnalysis}
\SetKwProg{Fn}{Function}{}{end}
\SetKwProg{Fp}{Procedure}{}{end}
\caption{\small Compiler Faults Isolation with \tool}
\label{algo:overall_algo}
\DontPrintSemicolon
\footnotesize
\KwIn{$BR$: Bug report.}
\KwOut{$SusFiles$: Suspicious compiler files.}

\BlankLine 
\tcc{Initial commit retrieval}
$BadRev, conf, prog \leftarrow$ \FnInfoRetrieve($BR$) \label{algo:info_retrieve} \;

\BlankLine 
\tcc{Rough range identification}
$release_{bad} \leftarrow$ \textbf{None} \\
\ForEach{$major \in \FnGetMajorReleases()$ \label{algo:bad_commit:start}}{
    \If{\FnCompile($major, conf, prog$) $==$ \textbf{Fail}}{
        \If{\FnDate($major$) $<$ \FnDate($release_{bad}$)}{
            $release_{bad} \leftarrow major$ \label{algo:bad_commit:end} \;

        }
    }
}
$release_{good} \leftarrow$ \FnGetPrecedingRelease($release_{bad}$) \label{algo:good_commit} \;
\BlankLine 
\tcc{Fine range identification}
\ForEach{$minor$ $\in$ \FnGetMinorReleases($release_{bad}$) \label{algo:minor:start}}{
    \If{\FnCompile($minor, conf, prog$) $==$ \textbf{Pass}}{
        \If{\FnDate($minor$) $>$ \FnDate($release_{good}$)}{
            $release_{good} \leftarrow minor$ \;
        }
    }
    \Else{
        \If{\FnDate($minor$) $<$ \FnDate($release_{bad}$)}{
            $release_{bad} \leftarrow minor$ \label{algo:minor:end} \;
        }
    }
}

$BadRev \leftarrow$ \FnGetReleaseCommit($release_{bad}$) \;
$GoodRev \leftarrow$ \FnGetReleaseCommit($release_{good}$) \;

\BlankLine
\tcc{Bug-inducing commit identification}
\While{$BadRev - GoodRev \neq 1$}{
    $rev \leftarrow$ \FnBisect($BadRev, GoodRev$) \;
    \If{\FnCompile($rev, conf, prog$) $==$ \textbf{Fail}}{
        $BadRev \leftarrow rev$ \;
    }
    \Else{
        $GoodRev \leftarrow rev$ \;
    }
}

\BlankLine 
\tcc{Step 5: Differential Analysis}
$SuspiciousFiles \leftarrow$ \FnDiffAnalysis($BadRev, GoodRev$) \;

\BlankLine  
\Return{$SuspiciousFiles$} \;

\end{algorithm}

\subsection{Illustrative Example}

In this subsection, we present a concrete example to demonstrate how \tool~operates. 
This example is derived from the debugging process of a real-world bug\footnote{\url{https://bugs.llvm.org/show_bug.cgi?id=33119}}.
Initially, \tool~retrieves the commit ID corresponding to the reported revision of the compiler that exhibits the bug.
Next, \tool~identifies the `good' and `bad' commits by examining the major and minor versions of the compiler.
Once the `good' and `bad' commits are determined, \tool~initiates the bisecting process using the command \texttt{git bisect start}, preparing the current working directory for bisecting between these commits.
\tool~then marks the known good and bad commits with \texttt{git bisect good <commit-hash>} and \texttt{git bisect bad <commit-hash>}, respectively.
Subsequently, git automatically checks out a commit that lies about halfway between the known good and bad commits. 
\tool~compiles the project and runs tests to determine if the bug manifests.
If the midpoint commit does not exhibit the bug (i.e., the bug is absent), it is labeled as `good' using `\texttt{git bisect good}'; otherwise, it is labeled as `bad' using `\texttt{git bisect bad}'.
This iterative process of testing and marking continues, successively halving the search range until it narrows down to zero revisions.
The last commit marked as `bad' is then confirmed as the bug-inducing commit. 
Finally, \tool~performs a differential analysis between the bug-inducing change and its preceding change to identify the code files modified by the commit.

\subsection{Implementation}

We implement \tool~as a practical automated compiler fault isolation tool. 
To achieve this, we utilize the git or svn version control system, which enables access to and extraction of the commit history of the compiler under investigation. 
By leveraging this resource, we trace the compiler's evolution over time, identifying specific commits that may have introduced faults. 
Furthermore, our implementation leverages pre-built binaries provided by developers for each release version of the compiler. 
Instead of recompiling the compiler for every version, we directly utilize these binaries to establish approximate ranges for identifying both the `bad commit' and the `good commit'.
This approach offers distinct advantages in computational efficiency and significantly expedites the fault isolation process. By avoiding the need to recompile the compiler for each version, we minimize computational overhead associated with isolation. Consequently, the time required to pinpoint the faulty commit is substantially reduced, facilitating timely resolution of compiler faults. 
Finally, we emphasize that our implementation only reflects one realistic instantiation of the workflow commonly adopted by developers. 
In practice, different compiler developers may employ a variety of scripts, heuristics, and strategies when identifying bug-inducing commits~\cite{llvmbisectdiscuss}.

\section{Experimental Settings}
\label{sec:settings}
\newcommand{\cis}{C}
\newcommand{\nis}{\overline{C}}

\subsection{Research Questions}
In this study, we aim to address the following research questions:

\begin{itemize}[leftmargin=*]
  \item \textbf{RQ1}: How does \tool~perform compared with existing compiler fault isolation techniques?
  \item \textbf{RQ2}: How does \tool~perform relative to representative SBFL-based techniques under different suspiciousness formulas?
  \item \textbf{RQ3}: Does \tool~complement state-of-the-art compiler fault isolation techniques?
\end{itemize}

\textit{RQ1} evaluates how \tool~compares with traditional SBFL-based techniques in terms of fault localization performance. The goal is to assess whether these established methods provide essential benefits over the straightforward, developer-aligned approach used by \tool~in accurately identifying faulty compiler files.

\textit{RQ2} builds on this comparison by examining whether a recent, state-of-the-art SBFL-based technique outperforms \tool~when different suspiciousness formulas are applied. This helps us better understand the sensitivity of fault localization performance to formula choice and the relative strengths of the underlying techniques.

\textit{RQ3} explores whether \tool~offers complementary benefits when used alongside SBFL-based techniques. Identifying such complementarity could inform future designs of hybrid approaches for compiler fault localization, offering practical guidance to researchers and practitioners.

\subsection{Environment and Hardware}

In order to ensure reliable and reproducible results, the experiments conducted in this study were carried out in a meticulously controlled environment. To establish a consistent and isolated experimental setting, we employed a Docker container, which offered a virtualized environment for executing the experiments. The Docker container was meticulously configured with a 20-core CPU and 120GB of memory, providing ample computational resources to effectively handle the experimental workload.
Regarding the choice of the underlying operating system, we opted for Ubuntu 14.04, a 64-bit variant, as the host operating system for the Docker container. This selection was based on its compatibility with the experimental setup and the availability of necessary software dependencies. Ubuntu 14.04 is widely recognized as a well-established and extensively used Linux distribution that is highly regarded for its stability and long-term support. By employing Ubuntu 14.04, we ensured the reliability and consistency of the experimental environment.
Furthermore, the utilization of Ubuntu 14.04 allows for the easy installation of various historical versions of GCC. This capability enables us to accurately determine the approximate range of bug-inducing commit, as most historical versions of GCC can be readily installed on Ubuntu 14.04. 

\subsection{Benchmark Selection}

The selection of appropriate benchmarks plays a crucial role in comparing the effectiveness of the existing SBFL-based techniques with the \tool~approach. 
Following recent work~\cite{ODFL-TSE}, we use a benchmark consisting of 60 real-world GCC bugs and 60 real-world LLVM bugs to evaluate \tool~against state-of-the-art SBFL-based techniques. 
Using this benchmark enables direct and fair comparisons with prior studies.
The 60 GCC bugs was originally collected by Chen et al.~\cite{chen2019compiler, chen2020enhanced} and were later used to evaluate LLM4CBI and \textsc{Odfl}~\cite{DBLP:journals/tse/TuZJYLJ24,ODFL-TSE}. 
For the LLVM benchmark, 20 bugs were initially collected by Chen et al.~\cite{chen2020enhanced}, while an additional 30 bugs were collected by Yang et al.~\cite{ODFL-TSE}.
The latter \textsc{Odfl} study relied on the ``-opt-bisect-limit'' feature to determine the failing optimization passes and this feature was introduced in LLVM version 267022 (April 22, 2016).\footnote{https://github.com/llvm/llvm-project/commit/f0f2792} To expand their evaluation, the authors re-collected 30 additional LLVM bugs, resulting in a total of 50 LLVM bugs. To maintain consistent scale of LLVM benchmark used in prior studies~\cite{chen2019compiler} and to further expand the evaluation set, we additionally collected 10 more LLVM bugs, yielding a total of 60 LLVM bugs in our benchmark. 
We select these 120 real-world compiler bugs due to the following reasons. First, all these bugs have been successfully addressed by developers, thus enabling us to identify faulty code based on bug-fixing information. 
Second, each bug in the benchmark was accompanied by essential information, including the lowest optimization level at which the bug was triggered, the corresponding test program that exposed the bug, the specific compiler version in which the bug was present, and the compiler files responsible for the manifestation of the bug (i.e., faulty files within compiler).
This selection of benchmarks served as the ground truth against which the effectiveness of investigated techniques needed to be evaluated. 
By utilizing such a benchmark, we aim to provide a comprehensive and objective assessment of the capabilities and performance of existing SBFL-based techniques in comparison to the simplified and straightforward \tool~technique. 
Furthermore, we chose GCC as our benchmark to enable a direct comparison between \tool~and \textsc{Odfl}~\cite{ODFL-TSE}, as \textsc{Odfl} was specifically implemented for GCC. 
However, it is important to note that \tool~is not limited to GCC but can also be applied to LLVM.

\subsection{Implementation and Parameters}

To ensure consistency with previous studies~\cite{chen2019compiler, chen2020enhanced, ODFL}, we implemented and evaluated existing SBFL-based techniques, including RecBi, DiWi, and \textsc{Odfl}. 
To collect coverage information of the compilers, we utilized Gcov~\cite{gcov}, specifically version 6.0.0. In cases of compatibility issues between GCC and Gcov 6.0.0, we use a compatible Gcov version.
The experiments were designed deterministically to enhance overall reliability. 
GCC was chosen as the target compiler due to its widespread usage and large number of fine-grained optimization options, facilitating \textsc{Odfl} implementation. 

In this study, we installed the required major and minor versions on our server. To use~\tool, developers need to install necessary compiler binaries for major and minor versions on their local machines or shared servers. However, if some binaries are unavailable, they can be installed by first downloading necessary support libraries, configuring the compiler, and building and installing the commit version (e.g., install gcc version with hashid 48a320a): 

\begin{lstlisting}[language=bash, basicstyle=\ttfamily\small]
$ git checkout 48a320a
$ ./contrib/download_prerequisites
$ mkdir buildgcc && cd buildgcc
$ ../gcc-git/configure --disable-multilib  
  --disable-bootstrap --enable-languages=c,c++ 
  --prefix=/usr/local/48a320a
$ make -j 40 && make install
\end{lstlisting}

\noindent We developed a Python-script to automate the process of dynamically installing necessary compiler versions on demand, without requiring pre-installation. Alternatively, online resources like the Compiler-Explorer\footnote{https://godbolt.org/} platform can be used if local installation is not desired. The Compiler-Explorer maintains a comprehensive collection of compiler versions, including latest trunk versions, and provides a REST-API for programmatically interacting with the platform and accessing required compiler binaries.

\subsection{Subject SBFL-based Techniques}

We compare \tool~with a set of advanced SBFL-based techniques, including DiWi~\cite{chen2019compiler}, RecBi~\cite{chen2020enhanced}, LLM4CBI~\cite{DBLP:journals/tse/TuZJYLJ24}, ETEM~\cite{ETEM-ASE24-hao}, \textsc{Odfl}~\cite{ODFL-TSE}, and HSFL~\cite{DBLP:journals/tse/WenCTWHHC21}. Among these, the first three are specifically designed for compiler fault isolation, while the last is a general-purpose SBFL technique. These techniques can be categorized into three groups. DiWi, RecBi, LLM4CBI, and ETEM are based on witness test programs, \textsc{Odfl} is based on adversarial configuration, and HSFL is based on historical spectra. 

\begin{itemize}[leftmargin=*]
  \item Witness test program-based SBFL: DiWi, RecBi, LLM4CBI, and ETEM generate passing test programs by mutating on original bug-triggering test case. DiWi applies local operation mutations and uses the Metropolis-Hastings algorithm to produce diverse program variants. RecBi leverages structural mutation along with reinforcement learning to guide the mutated process. ETEM generates diverse test programs using feature mutation operators. LLM4CBI employs Large Language Model (LLM) for generating witness test programs.
  \item Adversarial configuration-based SBFL: Rather than mutating test programs, \textsc{Odfl} crafts adversarial compilation configurations by selectively adjusting fine-grained optimization options. This strategy generates both passing and failing configurations, facilitating the collection of diverse coverage data and supporting effective fault isolation.  
  \item Historical spectrum-based SBFL: HSFL utilizes version history to aid fault localization. Its core insight is that program entities modified by bug-inducing commits, but rarely by subsequent non-inducing ones, are more likely to be faulty. HSFL first identifies bug-introducing commits, then constructs a historical spectrum-termed \textit{Histrum}, which is orthogonal to the conventional coverage-based spectrum used in SBFL. Although HSFL was not originally designed for compiler fault isolation, it can be readily adopted to this context. We re-implemented HSFL for compilers and included it as a baseline in our evaluation of~\tool.
\end{itemize}

All these SBFL-based techniques share the same high-level strategy: they use passing compilations to contrast against failing ones and isolate suspicious programs elements. Using both the failing and passing compilations, they compute suspiciousness scores for each program element and rank them accordingly. 

To quantify the suspiciousness of each statement, a standard Spectrum-Based Fault Localization (SBFL) formula is used.  
Specifically, the Ochiai formula~\cite{abreu2007accuracy} is adopted, as defined in Equation~\ref{formula:ochia}.
It considers the number of failing executions ($ef_{s}$) that cover a statement $s$, the number of passing executions ($ep_{s}$) that also cover $s$, and the number of failing executions that do not cover $s$ ($nf_{s}$):

\begin{equation}
  score(s) = \frac{ef_{s}}{\sqrt{(ef_{s}+nf_{s})(ef_{s}+ep_{s})} }\label{formula:ochia}
\end{equation}

\noindent Notably, HSFL adjusts these suspiciousness scores using historical modification data. Specifically, for each statement $s$ modified by a bug-inducing commit, HSFL computes a historical suspiciousness score as follow:

\begin{equation}
  Histrum(s) = \frac{induce(s)}{\sqrt{induce(s) + noninduce(s)}}
\end{equation}

\noindent Here, $induce(s)$ denotes the number of inducing commits that modify statement $s$. In compiler fault localization scenarios, we typically have a single known bug-inducing commit, so $induce(s) = 1$. The value $noninduce(s)$ represents the number of non-inducing commits that also modified $s$. 

HSFL integrates this historical spectrum with the conventional coverage-based spectrum to produce the final suspiciousness score. For compiler fault isolation, we derive the conventional spectrum by comparing the coverage between the bug-triggering optimization level (e.g., \texttt{-O3}) and a nearby non-buggy level (e.g., \texttt{-O2}). The final HSFL score is computed as:

\begin{equation}
  HSFL(s) = 
  \begin{cases}
    (1 - \alpha) \cdot SBFL(s), & \text{if } s \in \mathabxcal{A} \land s \notin S_{c} \\
    (1- \alpha) \cdot SBFL(s) + \alpha \cdot Histrum(s), & \text{if } s \in \mathabxcal{A} \land s \in S_{c} \\
    0, & \text{otherwise}
  \end{cases}
\end{equation}

\noindent 
Here, $\mathabxcal{A}$ is the set of statements executed during the failing execution, and $S_{c}$ is the sets of statements modified in the bug-inducing commit. 
Following prior work~\cite{DBLP:journals/tse/WenCTWHHC21}, we set $\alpha$ to 0.5 by default. 

Once suspicious scores are compute for each executed statement in the identified compiler source files, we aggregate them to derive a file-level suspiciousness score. In line with previous studies~\cite{chen2019compiler,chen2020enhanced,DBLP:journals/tse/TuZJYLJ24,ODFL-TSE}, we compute the file score as the average of statement scores within it. 
This is formalized in Equation~\ref{formula:avg}, where $n_{f}$ is the number of statements in file $f$ covered by failing configurations:

\begin{equation}
    S_{avg}(f)
    = \left. \sum_{i=1}^{n_f} score(s_i) \;\right/ n_f
    \label{formula:avg}
\end{equation}

\subsection{Measurements}

To evaluate the effectiveness of the compiler fault localization approaches, we adopt a set of widely used evaluation metrics, consistent with prior studies~\cite{chen2019compiler,chen2020enhanced,sohn2017fluccs,ODFL-TSE}. 
These metrics provide a comprehensive assessment of each approach's ability to accurately identify faulty components:

\begin{itemize}[leftmargin=*]
\item \textbf{Top-N:} This metric reports the number of bugs for which at least one faulty file is ranked within the top $N$ positions of the result list. We consider $N = 1$, $5$, $10$, and $20$. A higher value indicates better performance for this metric. 
\item \textbf{Mean First Rank (MFR):} MFR computes the average rank of the first faulty file in the result list for each bug. It emphasizes how early the top faulty element is located. Lower values indicate better performance.
\item \textbf{Mean Average Rank (MAR):} MAR measures the average rank of all faulty files for each bug, offering a broader perspective on the ranking quality. A lower MAR indicates more accurate localization across all faulty elements.
\end{itemize}

These metrics enable a thorough and multi-faceted evaluation of the techniques. Their widespread use in prior work ensures comparability and consistency with existing studies.  
It is important to note that \tool~only considers files modified in BIC as faulty. As a result, the MFR and MAR metrics, which assume the possibility of multiple faulty files, are not applicable to \tool. Therefore, we report only the Top-N metrics when evaluating \tool.

\section{Experimental Results}
\label{sec:evaluation}

This section presents the experimental results addressing RQ1 and RQ2. Specifically, we compare \tool~with existing SBFL-based compiler fault isolation techniques and assess the effectiveness of different SBFL formulae in terms of bug localization.

\subsection{\textit{RQ1: How does \tool~perform compared with existing compiler fault isolation techniques?}}\label{sec:RQ1}

To ensure a fair and conservative comparison that aligns with real-world developer practices, we report the performance of \tool~under a worst-case evaluation setting. 
Specifically, consider a bug-inducing commit that modifies multiple files, denoted as $[f_{1}, f_{2}, f_{3}, f_{4}, f_{5}, f_{6}]$, where only file $f_{1}$ is the actual buggy file. In our evaluation, we conservative assume that \tool~ranks $f_{1}$ at the lowest position among the modified files, i.e., the six position in this example. 
Consequently, this case is counted as a failure for metrics that consider only the top-ranked files (e.g., Top-1 and Top-5). 
This evaluation strategy penalizes \tool~as much as possible and thus avoid overestimating the effectiveness of the BIC-based approach. 
Consequently, the reported results represent a lower bound on the practical performance of \tool.

\begin{table}[t]
  \renewcommand{\arraystretch}{1.2}
  \setlength{\tabcolsep}{1pt}
  \caption{Compiler fault location effectiveness comparison}
  \label{tbl:comparison}
  \centering
  \small
  \begin{tabular}{c|l|rr|rr|rr|rr}
    \toprule
    \multirow{2}{*}{} & \multirow{2}{*}{Tools} & \multicolumn{2}{c|}{Top-1} & \multicolumn{2}{c|}{Top-5} & \multicolumn{2}{c|}{Top-10} & \multicolumn{2}{c}{Top-20} \\
    \cmidrule(lr){3-4} \cmidrule(lr){5-6} \cmidrule(lr){7-8} \cmidrule(lr){9-10}
    & & Value & $\Uparrow$ & Value & $\Uparrow$ & Value & $\Uparrow$ & Value & $\Uparrow$\\
    \midrule
    \multirow{6}{*}{GCC} 
    & DiWi & 5.6 & 275.0\% & 19.9 & 75.9\% & 31.3 & 11.8\% & 41.7 & -16.1\% \\
    & RecBi & 7.7 & 172.7\% & 23.9 & 46.4\% & 34.2 & 2.3\% & 42.5 & -17.7\% \\
    & LLM4CBI & 9.3 & 125.8\% & 24.9 & 40.6\% & 36.4 & -3.8\% & 44.8 & -21.9\% \\
    & HSFL & 19.0 & 10.5\% & 22.0 & 59.1\% & 30.0 & 16.7\% & 38.0 & -7.9\% \\
    & ETEM & 20.0 & 5.0\% & \textbf{38.0} & -7.9\% & \textbf{44.0} & -20.5\% & \textbf{54.0} & -35.2\% \\
    & \textsc{Odfl} & 20.0 & 16.7\% & 31.0 & 12.9\% & 42.0 & -16.7\% & 47.0 & -25.5\% \\
    & \textbf{\tool} & \textbf{21.0} & -- & 35.0 & -- & 35.0 & -- & 35.0 & -- \\
    \midrule
    \multirow{5}{*}{LLVM} 
    & HSFL  & 22.0 &  22.7\% & 23.0  &  47.8\% & 25.0  &  36.0\% & 30.0  &  13.3\% \\
    & ETEM & 7.0 &  285.7\% & 13.0 &  161.5\% & 21.0 &  61.9\% & 28.0 &   21.4\% \\ 
    & \textsc{Odfl} & 16.0 & 68.8\% & 27.0 & 25.9\% & \textbf{36.0} & -5.6\% & \textbf{47.0} & -27.7\% \\
    & \textbf{\tool} & \textbf{27.0} & -- & \textbf{34.0} & -- & 34.0 & -- & 34.0 & -- \\
    \bottomrule
  \end{tabular}
  \caption*{\footnotesize Columns ``$\Uparrow$'' show the improvement rate of~\tool~over each technique for the corresponding metric. Missing values are indicated by --.}
  \vspace{-1em}
\end{table}

\subsubsection{Accuracy}
Table~\ref{tbl:comparison} presents a comparative evaluation of the fault localization effectiveness of \tool~against existing SBFL-based techniques.
Rows three to nine correspond to the GCC benchmark, while the last six rows report results on the LLVM benchmark. 
The third, fifth, seventh, and ninth columns (\texttt{Value}) indicate the absolute number of bugs localized within the Top-1, Top-5, Top-10, and Top-20 ranked files, respectively. 
The adjacent columns ($\Uparrow$) show the relative improvement (or degradation) of \tool~with respect to each baseline on the corresponding metric. 

As shown in Table~\ref{tbl:comparison}, \tool~consistently outperforms all state-of-the-art SBFL-based techniques on the crucial Top-1 metric across both compiler benchmarks. To ensure a robust and fair comparison, and to avoid inconsistencies arising from differences in implementations or experimental environments, we adopt the best-performing GCC results reported in the original ETEM and \textsc{Odfl} studies~\cite{ODFL-TSE,ETEM-ASE24-hao}, since all approaches were evaluated on the same set of 60 bugs in the GCC benchmark. In particular, \textsc{Odfl} achieves its best performance under the DStar formula~\cite{wong2013dstar}. The results for DiWi, RecBi, and LLM4CBI correspond to the averages over 10 runs reported in the LLM4CBI paper~\cite{DBLP:journals/tse/TuZJYLJ24}. We re-ran HSFL on the same set of 60 bugs in the GCC benchmark. 

For the LLVM benchmark, the results of other techniques were obtained by re-running their evaluations on the same set of 60 faults. 
We observe that DiWi, RecBi, and LLM4CBI perform substantially worse than techniques such as \textsc{Odfl} and HSFL on the GCC benchmark. Moreover, prior work has shown that DiWi and RecBi performs significantly worse than \textsc{Odfl} on 50 out of the 60 LLVM bugs considered in this study~\cite{ODFL-TSE}. 
In addition, both DiWi, RecBi, and LLM4CBI incur considerable overhead due to extensive program mutation or frequent interactions with large language models, resulting in poor cost-effectiveness. Consequently, we do not re-evaluate these techniques on the new LLVM benchmark comprising 60 bugs.

\begin{itemize}[leftmargin=*]
    \item \textbf{GCC Benchmark:} Among the 60 bugs in the GCC benchmark, \tool~successfully localizes 21 faults within the Top-1 ranked file, outperforming DiWi (5.6), RecBi (7.7), LLM4CBI (9.3), HSFL (19.0), ETEM (20.0), and \textsc{Odfl} (20.0). This corresponds to Top-1 improvements of 275.0\%, 172.7\%, 125.8\%, 10.5\%, 5.0\%, and 5.0\%, respectively. 
    Although $ETEM$ and $\textsc{Odfl}$ achieve higher accuracy on the Top-5, Top-10, and Top-20 metrics, Top-1 localization is widely regarded as the most critical criterion in practice, as developers typically inspect only the highest-ranked files. Given its simplicity and closer alignment with real-world debugging workflows, \tool~offers a strong practical advantage despite its conservative evaluation.

    \item \textbf{LLVM Benchmark:} In contrast to the GCC results, \tool~demonstrates superior performance across multiple metrics on the LLVM benchmark, achieving the highest effectiveness in the Top-1 (27.0) and Top-5 (34.0) metrics. This corresponds to improvements ranging from 22.7\%--285.7\% for Top-1, and 25.9\%--161.5\% for Top-5 over existing techniques. 
    Given the practical importance of Top-1 and Top-5 localization in compiler debugging, these results further highlight the effectiveness of leveraging historical commit information over spectrum-based techniques.
\end{itemize}

\noindent\textbf{Summary of RQ1.}
Across both the GCC and LLVM benchmarks, \tool~demonstrates consistently competitive fault localization performance compared to state-of-the-art SBFL-based techniques, particularly on the Top-1 metric, which is practically the most relevant indicator.
Even under a deliberately conservative evaluation setting that penalizes multi-file bug-inducing commits, \tool~achieves the best Top-1 accuracy on both benchmarks and additionally attains leading performance on higher-rank metrics on LLVM.
These findings indicate that a simple BIC-based strategy can match or even outperform more sophisticated SBFL-based approaches in realistic compiler debugging scenarios.
Overall, the results of RQ1 provide empirical evidence that \tool~constitutes a practical, effective, and developer-aligned baseline.
This suggests that the anticipated progress in SBFL-based compiler fault isolation techniques may not have been achieved as initially envisioned. 
We therefore recommend adopting \tool~as a strong reference approach for future research on compiler fault localization.

\subsection{\textit{RQ2: How does \tool~perform relative to representative SBFL-based techniques under different suspiciousness formulas?}}

\begin{table}[t]
	\centering
	\caption{The six adapted SBFL formulae}
	\label{formula}
	\normalsize
	\setlength\tabcolsep{9pt} %
	\def\arraystretch{1.5}
	\begin{tabular}{ll}
		\toprule
		Name & Formula \\ \hline

		\textbf{Ochiai} \cite{abreu2007accuracy} & $\frac{ef_{s}}{\sqrt{(ef_{s}+nf_{s})(ef_{s}+ep_{s})} }$ \\

		\textbf{Tarantula} \cite{jones2005empirical} & $\frac{ef_{s}/totalf_s}{ef_s/totalf_s + ep_s/totalp_s}$ \\
  
        \textbf{Ochiai2} \cite{naish2011model}    & $\frac{ef_{s}np_{s}}{\sqrt{(ef_{s}+ep_{s})(nf_{s}+np_{s})(ef_{s}+np_{s})(nf_{s}+ep_{s})}} $\\

        \textbf{Op2} \cite{naish2011model}       & $ef_{s} - \frac{ep_{s}}{ (totalp_{s} + 1)}$ \\

        \textbf{Barinel} \cite{abreu2009spectrum}   & $1 - \frac{ep_{s}}{ep_{s} + ef_{s}} $\\
        
		\textbf{DStar} \cite{wong2013dstar}     & $\frac{ef_{s}^\star}{ep_{s} + nf_{s}} $ \\
		\bottomrule
		\multicolumn{2}{l}{\scriptsize We used $\star = 2$, the most thoroughly explored value \cite{pearson2017evaluating}}
	\end{tabular}
\vspace{-1em}   %
\end{table}

\begin{table}[t]
  \renewcommand{\arraystretch}{1.2}
  \setlength{\tabcolsep}{1.5pt}
  \caption{Performance comparison for \tool~and \textsc{Odfl} in terms of different formulas}
  \label{tbl:comparison-formulae}
  \centering
  \footnotesize
  \begin{tabular}{l|r|cccc||cccc}
      \toprule
      \multirow{2}{*}{Tool} & \multirow{2}{*}{Formula} & \multicolumn{4}{c||}{GCC} & \multicolumn{4}{c}{LLVM} \\
      \cmidrule{3-6} \cmidrule{7-10}
      & & \textbf{Top-1} & \textbf{Top-5} & \textbf{Top-10} & \textbf{Top-20} & \textbf{Top-1} & \textbf{Top-5} & \textbf{Top-10} & \textbf{Top-20} \\
      \midrule
      \multirow{6}{*}{\textsc{Odfl}} & Ochiai   & 18 & 32 & 41 & 47 & 16 & 27 & 36 & 47 \\
      & Tarantula  & 14 & 32 & 39 & \textbf{49} & 16 & 29 & \textbf{39} & \textbf{50} \\
      & Ochiai2    & 14 & 26 & 36 & 46 &  9 &  23 & 25  & 21 \\
      & Op2       & 16 & 27 & 40 & 46 & 9  & 22 & 28 & 36 \\
      & Barinel   & 18 & 32 & 38 & 48 & 13 & 28 & 35 & \textbf{50} \\
      & DStar     & 20 & 31 & \textbf{42} & 47 & 15 & 25 & 34 & 45 \\
      \midrule
      \tool & - & \textbf{21} & \textbf{35} & 35 & 35 & \textbf{27} & \textbf{34} & 34 & 34 \\
      \bottomrule
  \end{tabular}
  \vspace{-1.0em}
\end{table}

Over the past two decades, Spectrum-Based Fault Localization (SBFL) has made significant progress, resulting in the development of a wide range of formulas for computing suspiciousness scores~\cite{wong2016survey,tosem2013xie}.
In Section~\ref{sec:RQ1}, we a adopt specific suspiciousness formula to compute statement-level scores for SBFL-based compiler fault localization techniques. 
However, it remains unclear whether \tool~can achieve comparable or even superior performance relative to existing SBFL-based techniques when different suspiciousness formulas are employed. 

To address this question and enable a comprehensive evaluation of \tool, we systematically compare \tool~with \textsc{Odfl} across a diverse set of SBFL formulas. 
We focus exclusively on \textsc{Odfl} for two main reasons. 
First, as demonstrated in RQ1, \textsc{Odfl} emerges as the most cost-effective SBFL-based compiler fault localization technique: it incurs relatively low computational overhead while achieving competitive performance. 
Second, \textsc{Odfl} represents the most effective compiler fault localization approach that relies solely on SBFL, without incorporating any commit-based analysis.
This makes \textsc{Odfl} an ideal representative for reassessing the effectiveness of pure SBFL-based techniques by contrasting them with \tool, a simple strategy that leverages commit-based information and aligns closely with real-world developer practices.

For this research question, we adopt a diverse SBFL formulae variations described in a previous study~\cite{wong2016survey} to re-evaluate \textsc{Odfl}. 
Table~\ref{formula} provides an overview of the SBFL formulas investigated in this study. 
In these formulas, $ef_s$, $nf_s$, and $ep_s$ are defined in the same way as explained in Formula~\ref{formula:ochia}, while $np_s$ denotes the number of passing cases that do not execute the statement $s$, and $totalf_s$ and $totalp_s$ denote the total number of failing cases and passing cases for statement $s$, respectively. 

Table~\ref{tbl:comparison-formulae} presents the results of this comparison, showcasing the performance of \textsc{Odfl} under different suspiciousness formulas. 
Each row in Table~\ref{tbl:comparison-formulae} corresponds to a specific formula and presents the associated localization results. 
As shown in Table~\ref{tbl:comparison-formulae}, the performance of \textsc{Odfl} varies substantially across different formulas, particularly with respect to the Top-1 and Top-5 metrics.
On the GCC benchmark, the number of faults localized by \textsc{Odfl} under the Top-1 metric ranges from 14 to 20, while under the Top-5 metric it ranges from 26 to 32.
Despite this variability, \tool~consistently outperforms \textsc{Odfl} in terms of both Top-1 and Top-5 performance.
Similar trends are observed on the LLVM benchmark.
In summary, our results indicate that the SBFL-based compiler fault localization technique \textsc{Odfl} is sensitive to the choice of suspiciousness formula and exhibits notable performance instability.
In contrast, \tool~consistently outperforms \textsc{Odfl} on the two most critical metrics, Top-1 and Top-5, demonstrating stronger robustness across different SBFL formulas.

\subsection{\textit{RQ3: Does \tool~complement state-of-the-art compiler fault isolation techniques?}}

\begin{figure}[t]
    \centering
    \begin{subfigure}{0.23\textwidth}
        \centering
        \includegraphics[width=0.9\linewidth]{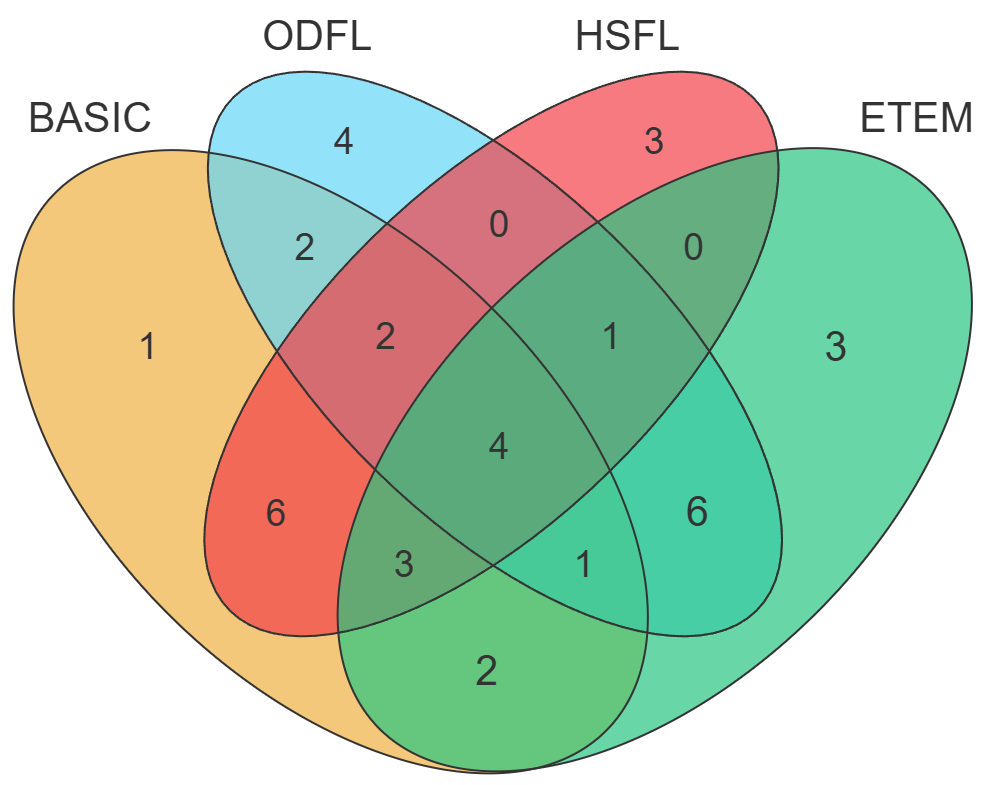}
        \caption{GCC}
        \label{fig:venn-compare1}
    \end{subfigure}
    \hfill
    \begin{subfigure}{0.23\textwidth}
        \centering
        \includegraphics[width=0.9\linewidth]{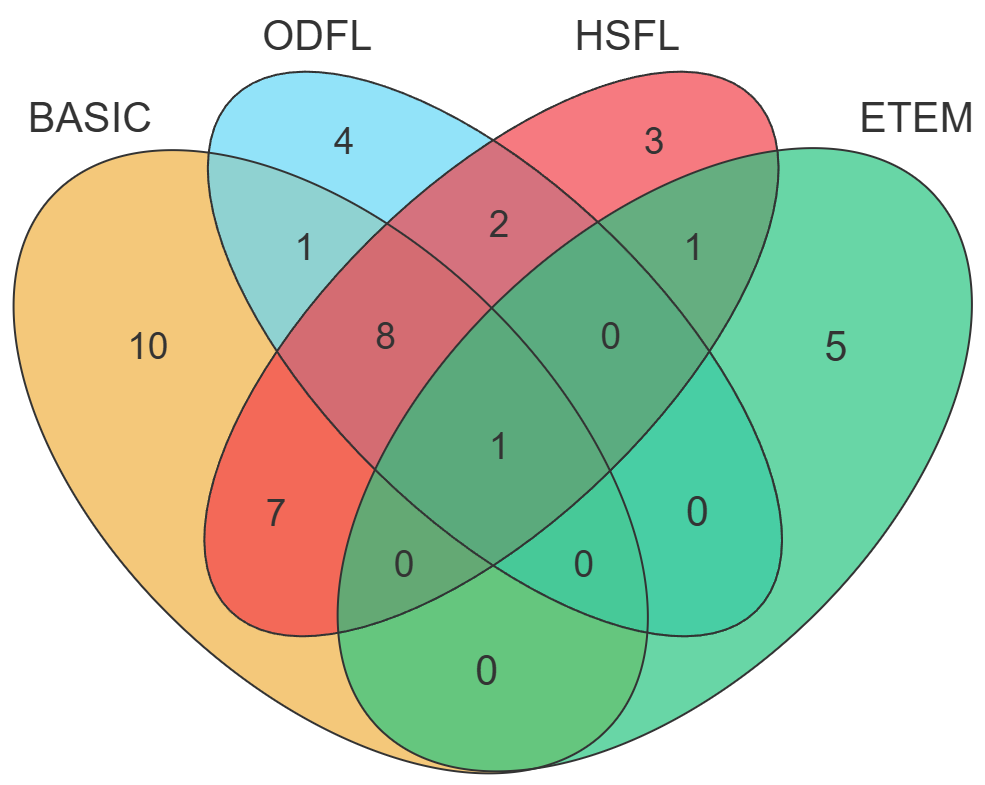}
        \caption{LLVM}
        \label{fig:venn-compare2}
    \end{subfigure}
\caption{Distribution of correct localization achieved by \tool, HSFL, ETEM and \textsc{Odfl} in terms of Top-1. \label{fig:venn}}
\vspace{-1em}
\end{figure}

To assess whether \tool~is not only effective in isolation but also complementary to state-of-the-art SBFL techniques, we analyze the overlap of faults correctly localized by \tool, HSFL, ETEM, and \textsc{Odfl} under the Top-1 metric. We select HSFL, ETEM, and \textsc{Odfl} as baselines due to their strong performance in earlier experiments in either GCC or LLVM. The results are illustrated in Figure~\ref{fig:venn}.

On the GCC benchmark comprising 60 bugs (Figure~\ref{fig:venn-compare1}), \tool~correctly localizes 21 faults at Top-1. Among these, 12 are not localized by \textsc{Odfl}, 6 are not localized by HSFL, and 11 are not detected by ETEM. Conversely, \textsc{Odfl}, HSFL, and ETEM correctly localize 11, 4, and 10 faults, respectively, that are not localized by \tool. 

On the LLVM benchmark comprising 60 bugs, \tool~correctly localize 27 bugs at Top-1. Notably, among these, 17 faults are not localized by \textsc{Odfl}, 11 are not localized by HSFL, and 26 are not localized by ETEM, indicating that \tool~identifies a substantial number of faults missed by existing SBFL-based techniques. 
In contrast, \textsc{Odfl}, HSFL, and ETEM together correctly localize only 6 faults that are not localized by \tool.

Overall, these results demonstrate that \tool~and SBFL-based techniques localize partially disjoint sets of faults, suggesting that \tool~provides complementary strengths. Consequently, effectively integrating \tool~with existing SBFL approaches has the potential to further improve overall compiler fault localization performance.

\section{Discussion}
\label{sec:validity}

In this section, we discuss the efficiency of \tool~and the threats to validity of this study.

\subsection{Efficiency}

The efficiency of \tool~is crucial for its practical viability in software debugging. Our analysis indicates that \tool~requires approximately 1.5 hours on average to identify bug-inducing commits via binary search, positioning it as a competitive alternative to SBFL techniques such as DiWi and RecBi.
The primary cost of \tool~arises from setting up and installing various compiler versions during the binary search, which typically involves about six iterations. Each commit version requires roughly 800 seconds for compilation and installation, and utilizing a parallel build configuration of 40 jobs optimizes this process.
In contrast, DiWi and RecBi take around five hours to generate the necessary witness test programs, which can be burdensome in fast-paced development environments where rapid bug identification is essential.
Although \tool~is more time-consuming than the \textsc{Odfl} approach (which averages about 0.6 hours for fine-grained optimization searches), it remains less costly than other SBFL techniques, further demonstrating its practicality.

\subsection{Threat to Validity}

\mypara{Benchmark Selection:} 
One potential threat arises from the possibility of flawed sample selection, as different samples may yield varying results. However, for direct comparison with three existing studies, a limited set of 60 defects in the GCC compiler and 60 defects in the LLVM compiler was selected as the sample. To ensure more robust conclusions, it is recommended to further analyze and compare the existing SBFL-based method with the \tool~using a larger set of defect samples.

\mypara{Quantitative Indicators:} Our evaluation relies on common Top-$k$ metrics (Top-1, Top-5, Top-10, and Top-20). While additional measures such as MAR and MFR can capture average ranking performance, they are not applicable to \tool, as it only considers files affected by a bug-introducing commit rather than the entire compiler. In practice, Top-1 and Top-5 are the most actionable to developers, so we expect this limitation to have minimal impact on the study's conclusions.

\mypara{Method of Comparison:}
Another threat concerns the comparison strategy. SBFL-based methods may assign identical suspiciousness scores to multiple files. To ensure a fair comparison, we report their best-case ranking by placing the true faulty file first among ties. For \tool, we report the worst-case ranking by placing the faulty file last when multiple files are touched by the bug-introducing commit. This conservative setup provides a more reliable assessment of the relative performance of the SBFL-based method and \tool.

\section{Related Work}
\label{sec:related-work}

\mypara{{Automated Fault Localization.}}
Debugging is essential for helping developers understand and resolve software defects~\cite{10.1145/3763093,DBLP:conf/issre/WangWZXZZ23,DBLP:journals/smr/ZhangLLJL24,DBLP:conf/sigsoft/LiM00C020,DBLP:journals/chinaf/SongXX24}. 
Within this process, automated fault localization plays a key role by narrowing down the parts of a program that are most likely to contain the root cause of a failure. 
Existing techniques include spectrum-based \cite{abreu2009spectrum, tang2017accuracy,wen2019historical}, slicing-based \cite{agrawal1993debugging, zhang2005experimental}, mutation-based \cite{moon2014ask, hong2015mutation, papadakis2015metallaxis}, and model-based approaches \cite{jung2018combining, wen2016locus}.
However, the structural and semantic complexity of compilers often limits the applicability and precision of these general-purpose techniques. 
To address these challenges, Chen et al. introduced DiWi \cite{chen2019compiler} and RecBi \cite{chen2020enhanced}, which leverage witness programs to aid in bug isolation, though both techniques still face constraints in effectiveness and efficiency. Tu et al. further extended this direction with LLM4CBI~\cite{DBLP:journals/tse/TuZJYLJ24}, which uses large language models to generate witness test programs. In parallel, Yang et al. proposed \textsc{Odfl}~\cite{ODFL,ODFL-TSE}, which derives passing and failing executions by adjusting compiler optimization configurations. 
Recent work by Zhou et al. has explored the use of historical bug-inducing commits for compiler bug deduplication~\cite{ZhouXuSun2025}, demonstrating the potential of BIC-based analysis for debugging compiler issues.
In this study, we introduce \tool, a practical bisection-driven strategy tailored for compiler fault localization. 
\tool~directly identifies the BIC and treats all modified files as potential suspects, aligning closely with developer practices. 
Our experiments show that \tool~performs comparably to, and often surpasses, state-of-the-art spectrum-based techniques in isolating compiler faults.

\mypara{Bug-inducing Commit Identification.}
Identifying bug-inducing commits is essential for debugging and fixing software issues~\cite{DBLP:conf/sigsoft/WenWLTXCS19}. Existing approaches fall broadly into three categories~\cite{DBLP:conf/icse/AnHKY23}. 
(1) Static methods analyze commit histories or bug reports without executing test cases. The canonical example is the SZZ algorithm and its variants~\cite{DBLP:journals/sigsoft/SliwerskiZZ05}, which trace from a bug-fixing commit to earlier commits that modified relevant elements.
(2) Information-retrieval-based approaches treat the task as a document-retrieval problem by matching bug reports with commit logs~\cite{DBLP:conf/kbse/WenWC16}.
(3) Dynamic methods employ execution information along with bug-exposing test inputs~\cite{DBLP:conf/sigsoft/AnY21, DBLP:conf/icse/AnHKY23}, using coverage-based analysis to identify the most suspicious commits.
In this work, we employ a binary-search-based strategy across the commit history to determine which revision introduces the observed bug. Despite is conceptual simplicity, bisection provides sufficiently precise BIC identification for our compiler fault isolation. 
Building on this foundation, \tool~integrates BIC-driven bisection into a specialized fault-localization framework, enabling a systematic comparison with SBFL-based techniques.

\section{Conclusion}
\label{sec:conclusion}

This study aims to rigorously evaluate the effectiveness of current spectrum-based compiler fault isolation techniques, particularly in comparison to \tool, a straightforward strategy commonly adopted by developers in practice.
Remarkably, \tool~demonstrates to be competitive with, and in some instances, superior to, existing SBFL-based techniques in terms of accuracy.
Such findings challenge the supposed advancements attributed to SBFL methodologies, especially their ability to accurately prioritize suspect files.
Given these insights, it is imperative for future research to include \tool~as a baseline in the evaluation of new compiler fault isolation strategies.
This will ensure a more rigorous assessment of novel techniques against established methods. The artifact of this work is publicly available at: \textbf{\url{https://doi.org/10.5281/zenodo.15030695}}.

\section*{Acknowledgements}

We thank the anonymous reviewers for their valuable feedback. We would like to express our appreciation to the authors of DiWi, RecBi, LLM4CBI, and \textsc{Odfl} for generously sharing their tools and data. Qingyang Li, Maolin Sun, and Yuming Zhou are the corresponding authors. This work was supported in part by the National Natural Science Foundation of China (Grants 624B2067 and 62572226), the Jiangsu Natural Science Foundation (Grant BK20231402), the Innovation Project of NJU-Huawei Joint Innovation Lab (Grant TC20230202021-2024-03), and the Collaborative Innovation Center of Novel Software Technology and Industrialization.

\balance
\bibliographystyle{ACM-Reference-Format}
\bibliography{full}

\end{document}